\documentclass[%
 reprint,                                         
 amsmath,amssymb,
 aps,    prc
]{revtex4-2}

\usepackage{dcolumn}
\usepackage{bm}

\usepackage{multirow}

\begin{document}

\title{Light kaonic atoms: from "corrected" to "summed up" Deser formula}

\author{N.V. Shevchenko}%
 \email{shevchenko@ujf.cas.cz}
\affiliation{%
 Nuclear Physics Institute, 25068 \v{R}e\v{z}, Czech Republic
}%

\date{\today}

\begin{abstract}
Accuracy of ''corrected Deser'' and ''summed up Deser'' formulas
was checked for the $K^- p$ and $K^- d$ systems. It was found that 
the last one is much more accurate and should be used for connection
the $1s$ level shift to the corresponding scattering length.
\end{abstract}


\maketitle

Interaction of antikaons with nucleons is the basic input for studying
quasi-bound states in exotic nuclei containing antikaons. Experimental data, which
can be used for construction of a model of $\bar{K}N$ interaction,
are not too rich. Some of the data are old and not very accurate,
such as $K^- p$ cross-sections. They can be reproduced by an antikaon-nucleon
strong potential directly, in a model independent way. Reproduction of
others cannot be used for parameter fitting directly, it needs some additional
model-dependent assumptions. It is the case of $K^- p$ correlation functions
recently measured by ALICE at CERN.

The most promising is kaonic hydrogen, which is an atom consisting of $K^-$ and
proton. It's $1s$ level is shifted in respect to the pure Coulomb value due to strong
interaction between the particles.  In principle, the shift and the width of kaonic hydrogen
is the data, which can be calculated directly using the strong plus Coulomb interactions.
However, it is much easier to calculate $K^- p$ scattering length using only strong
$\bar{K}N$ interaction, and then use some formula, which connects the scattering
length to the $1s$ level shift. Such formula was suggested by Deser for pion-nucleon
system \cite{Deser}.

Later the authors of \cite{corDeser} suggested a formula, which is more
accurate for the case when the meson is heavier, like an antikaon.
This ''corrected Deser'' formula is widely used nowadays in a form
\begin{equation}
\label{correctedDeser}
\Delta E_{K^-p}^{cD} - i \frac{\Gamma_{K^-p}^{cD}}{2} = \\
- 2 \alpha^3 \mu^2 a_{K^- p} [1-2 \alpha \mu a_{K^- p} (\ln{\alpha} -1)]
\end{equation}
by experimentalists and by theorists, who are not able to calculate $1s$ level of
kaonic hydrogen directly. It was shown in \cite{ourCrit,otherCrit}, that the accuracy
of the formula is about $10\%$ for the two-body $K^- p$ system and much worse
for the three-body $K^- d$ system.

However, the series, which leads to the corrected Deser formula in \cite{corDeser} 
can be summed up. The result is a formula, which I will call "summed up Deser'' formula
\begin{equation}
\label{summedDeser}
\Delta E_{K^-p}^{sD} - i \, \frac{\Gamma_{K^-p}^{sD}}{2} =
- 2 \alpha^3 \mu^2 a_{K^- p} / [1+2 \alpha \mu a_{K^- p} (\ln{\alpha} -1)].
\end{equation}
To the best of my knowledge, first it was wrote down in \cite{sumDeser}.
The authors of \cite{sumDeser} introduce it in a footnote: "In the case of kaonic atoms,
higher-order Coulomb corrections may turn out to be not completely negligible numerically
\dots This issue is, however, relatively easy to cure since the large contribution comes
from an iteration of a particular diagram to all orders. Replacing the factor
$1 - 2 \mu \alpha (\ln{\alpha} - 1) a_p$ by $(1 + 2 \mu \alpha (\ln{\alpha} - 1) a_p)^{-1}$
already captures the bulk of the effect. We shall not further elaborate on this issue."

Such summed up formula should be more accurate than the first two terms
of the series. The question is, how accurate is it, especially for the three-body
system.

Knowledge of the accuracy of the approximate Deser-type formulas is useful not
only for theorists, but for experimentalists as well. Quite a few experiments plan
or performing measurements of antikaon-nucleon or -nucleus systems. In particular,
SIDDHARTA-2~\cite{SIDDHARTA2} or J-PARK E57~\cite{JPARK_E57} experiments
aim to measure $1s$ level shift in kaonic deuterium.

Authors of \cite{upomin1,upomin2} mentioned in passing "summed up" Deser formula
as "substantially improving the agreement" or "reproducing the level shifts considerably
better". Accuracy of "summed up" Deser formula was checked
in \cite{checkSuDeser}, where for the two-body $K^- p$ system it was found to be much higher
than for the "corrected" Deser formula. The results for the three-body $K^- d$
system, however, are not so clear. The problem is that $1s$ level shift and width of kaonic
deuterium were evaluated in \cite{checkSuDeser} from variational calculation with
energy-independent $\bar{K}N$ potential, while $K^- d$ scattering length, necessary
for the approximate formulas, was calculated using the fixed-center approximation.
Therefore, it is not absolutely clear, what is the accuracy of the approximate
formulas themselves, and what are the errors introduced by the approximations.

In previous years we performed a series of calculations of diffferent
states and reactions in $\bar{K}NN$ and $\bar{K}\bar{K}N$ systems
 \cite{myReview}. The calculations were performed using dynamically
exact Faddeev-type AGS equations. As an input we used three
different versions of the $\bar{K}N$ interaction, constructed by us.
The potentials reproduce low-energy $K^- p$ experimental data including
$1s$ level shift and width of kaonic hydrogen. In contrast to most of
the authors of $\bar{K}N$ interaction models, we calculated energy of
the $1s$ level and its width directly by solving Lippmann-Schwinger
equation with strong plus Coulomb potentials. The scattering lengths,
given by the strong antikaon-nucleon potential, were also calculated, so that
we can compare the directly calculated characteristics of kaonic hydrogen
with those given by approximate formulas.

In addition, $1s$ shift caused by strong interaction in kaonic deuterium
was evaluated using Faddeev-type three-body equations with directly
included strong plus Coulomb interactions in \cite{Revai_Kdeu}. Since
$K^- d$ scattering length was calculated before as well, it is a
unique possibility to check the approximate formulas, connecting scattering
length with characteristics of kaonic atoms for two- and three-body systems,
comparing them with the exact results.

Three our antikaon-nucleon potentials are: two phenomenological potentials
having one-  $V_{\bar{K}N}^{\rm 1,SIDD}$  or two-pole $V_{\bar{K}N}^{\rm 2,SIDD}$
structure of $\Lambda(1405)$ resonance \cite{my_Kp_SIDD} , and a chiral potential
$V_{\bar{K}N}^{\rm Chiral}$ \cite{chiralKdlength}. All three potentials reproduce
the most recent and most accurate result of the $1s$ level shift of kaonic
hydrogen measurement performed by SIDDHARTA collaboration \cite{SIDDHARTA}.

The $K^- p$ scattering lengths of the three antikaon-nucleon potentials 
\cite{my_Kp_SIDD,chiralKdlength}, reproducing SIDDHARTA data are:
\begin{eqnarray}       
\label{aKpSIDD1}
&{}& a_{K^- p}^{\rm 1,SIDD} = -0.76 + i \, 0.89 \, {\rm fm} \\
\label{aKpSIDD2}
&{}& a_{K^- p}^{\rm 2,SIDD} = -0.74 + i \, 0.90  \, {\rm fm} \\
\label{aKpchiral}
&{}& a_{K^- p}^{\rm Chiral}  = -0.77 + i \, 0.84  \, {\rm fm.}
\end{eqnarray}
I also used our previously constructed phenomenological potentials from \cite{my_Kp_KEK},
which do not reproduce SIDDHARTA data, but can be useful for the checks of the
approximate formulas. The $K^- p$ scattering lengths given by the one-pole
$V_{\bar{K}N}^{\rm 1,KEK}$ and the two-pole $V_{\bar{K}N}^{\rm 1,KEK}$ potentials are:
\begin{eqnarray}
\label{aKpKEK1}
&{}& a_{K^- p}^{\rm 1,KEK} = -1.00 + i \, 0.68 \, {\rm fm} \\
\label{aKpKEK2}
&{}& a_{K^- p}^{\rm 2,KEK} = -0.96 + i \, 0.80 \, {\rm fm}.
\end{eqnarray}

\begin{table}
\caption{$1s$ level shift $\Delta E_{1s}^{K^- p}$ (eV) and width $\Gamma_{1s}^{K^- p}$ (eV)
of kaonic hydrogen calculated using the "corrected Deser" and "summed up Deser" formulas
together with the exact results.}
\begin{center}
\begin{tabular}{ccccccc}
    \hline \noalign{\smallskip}
    & \multicolumn{2}{c}{Corrected} & \multicolumn{2}{c}{Summed up} & \multicolumn{2}{c}{Exact} \\
    & \multicolumn{2}{c}{   Deser} & \multicolumn{2}{c}{    Deser} & \multicolumn{2}{c}{ } \\
    \hline \noalign{\smallskip}
    & $\Delta E_{1s}^{K^- p}$  & $\Gamma_{1s}^{K^- p}$ 
    & $\Delta E_{1s}^{K^- p}$  & $\Gamma_{1s}^{K^- p}$
    & $\Delta E_{1s}^{K^- p}$  & $\Gamma_{1s}^{K^- p}$ \\
    \hline \noalign{\smallskip}
$V_{\bar{K}N}^{\rm 1,SIDD}$ &  -328 & 579 & -318 & 593 & -313 & 597 \\[1mm]
$V_{\bar{K}N}^{\rm 2,SIDD}$ &  -322 & 589 & -312 & 603 & -308 & 602 \\[1mm]
$V_{\bar{K}N}^{\rm Chiral}$ &  -326 & 544 & -318 & 559 & -313 & 561 \\[1mm]
    \hline \noalign{\smallskip}
$V_{\bar{K}N}^{\rm 1,KEK}$ &  -383 & 404 & -381 & 429 & -377 & 434 \\[1mm]
$V_{\bar{K}N}^{\rm 2,KEK}$ &  -381 & 482 & -376 & 509 & -373 & 514 \\
    \hline
\end{tabular}
\end{center}
\label{Kp_tab}
\end{table}

The $1s$ level shifts and widths of kaonic hydrogen calculated using the "corrected Deser"
Eq.({\ref{correctedDeser}}) and "summed up Deser" Eq.(\ref{summedDeser}) formulas
are presented in Table \ref{Kp_tab} together with the exact results. It is seen that the "summed up
Deser" formula lead to much more accurate $\Delta E_{1s}^{K^- p}$ (eV) and
 $\Gamma_{1s}^{K^- p}$ values corresponding to the $K^- p$
scattering lengths Eqs.(\ref{aKpSIDD1},\ref{aKpSIDD2},\ref{aKpchiral}) and
Eq.(\ref{aKpKEK1},\ref{aKpKEK2}) than the "corrected Deser" formula.
The accuracy of the "summed up Deser" formula 
variates within $0.8-1.6 \%$ for the shift and $0.2-1.1 \%$ for the width.

The three potentials reproducing SIDDHARTA experimental data:
phenomenological $V_{\bar{K}N}^{\rm 1/2,SIDD}$ \cite{my_Kp_SIDD} and
chiral potential $V_{\bar{K}N}^{\rm Chiral}$  \cite{chiralKdlength}, -
were used in the three-body calculations of the $\bar{K}NN$ system with spin $1$.
The $K^- d$ scattering lengths were evaluated by solving the dynamically exact
Faddeev-type AGS equations with strong $\bar{K}N$ potentials only:
\begin{eqnarray}                 
\label{aKdSIDD1}
&{}& a_{K^- d}^{\rm 1,SIDD} = -1.49 + i 1.24 \, {\rm fm} \\
\label{aKdSIDD2}
&{}& a_{K^- d}^{\rm 2,SIDD} = -1.51 + i 1.25  \, {\rm fm} \\
\label{aKdchiral}
&{}& a_{K^- d}^{\rm Chiral}  = -1.59 + i 1.32  \, {\rm fm.}
\end{eqnarray}

Dynamically exact calculation of the kaonic deuterium characteristics were performed
using Faddeev-type equations with directly included Coulomb interaction in \cite{Revai_Kdeu}.
The same three strong antikaon-nucleon potentials were used there. The $1s$ level shifts and widths
of the three-body kaonic atom corresponding to the scattering lengths of $K^- d$ system
Eq.(\ref{aKdSIDD1},\ref{aKdSIDD2},\ref{aKdchiral})
are presented in Table \ref{Kd_tab}. We used the same Eq.(\ref{correctedDeser})
and Eq.(\ref{summedDeser}) for the "corrected" and "summed up Deser" formulas as
for the two-body $K^- p$ case with $\mu$ being the $K^- d$ reduced mass.

The results of one more method of kaonic deuterium characteristics calculations are
denoted in Table \ref{Kd_tab} as "$K^- d$ optical potential". To obtain them we:
calculated low-energy $K^- d$ scattering amplitudes using dynamically exact three-body
AGS equations, constructed an optical $K^- - d$ two-body potential, reproducing the
three-body amplitudes, and evaluated the $1s$ level shift and width using two-body
Lippmann-Schwinger equations with the optical $K^- - d$ and Coulomb potentials
\cite{my_Kp_SIDD}.

\begin{table*}
\caption{$1s$ level shift $\Delta E_{1s}^{K^- d}$ (eV) and width $\Gamma_{1s}^{K^- d}$ (eV)
of kaonic deuterium calculated using the "corrected Deser", "summed up Deser" formulas 
and using $K^ - d$ optical potentials together with the exact results.}
\begin{center}
\begin{tabular}{ccccccccc}
    \hline \noalign{\smallskip}
    & \multicolumn{2}{c}{Corrected} & \multicolumn{2}{c}{Summed up} & \multicolumn{2}{c}{Exact} 
    & \multicolumn{2}{c}{$K^- d$ optical} \\
    & \multicolumn{2}{c}{   Deser} & \multicolumn{2}{c}{    Deser} & \multicolumn{2}{c}{} 
    & \multicolumn{2}{c}{potential} \\
    \hline \noalign{\smallskip}
    & $\Delta E_{1s}^{K^- p}$  & $\Gamma_{1s}^{K^- p}$ 
    & $\Delta E_{1s}^{K^- p}$  & $\Gamma_{1s}^{K^- p}$
    & $\Delta E_{1s}^{K^- p}$  & $\Gamma_{1s}^{K^- p}$
    & $\Delta E_{1s}^{K^- p}$  & $\Gamma_{1s}^{K^- p}$ \\
    \hline \noalign{\smallskip}
$V_{\bar{K}N}^{\rm 1,SIDD}$ &  -826 & 731 & -792 & 921 & -767 & 928 & -785 & 1018 \\[1mm]
$V_{\bar{K}N}^{\rm 2,SIDD}$ &  -835 & 727 & -800 & 923 & -782 & 938 & -797 & 1025 \\[1mm]
$V_{\bar{K}N}^{\rm Chiral}$ &  -876 & 724 & -836 & 951 & -835 & 1004 & -828 & 1055 \\[1mm]
    \hline
\end{tabular}
\end{center}
\label{Kd_tab}
\end{table*}

It is seen from Table \ref{Kd_tab} that in this three-body case "summed up Deser" result is
slightly worse in reproducing the exact results: $0.1-3.3 \%$ for the shift and $0.8-5.3 \%$
for the width, - but is much better than the results of "corrected Deser" formula.
As for the "$K^- d$ optical potential" method, it gives comparable accuracy for the $1s$ shift
($0.8 - 2.4 \%$), but is worse for the width ($5.1 - 9.7 \%$) than the "summed up Deser".

Keeping all the above in mind, I suggest to use "summed up" Deser formula Eq.(\ref{summedDeser})
for evaluation $1s$ level shift and width of kaonic hydrogen and kaonic deuterium from $K^- p$ and
$K^- d$ scattering length correspondingly. Its accuracy is better than $2 \%$ for the two-body
$K^- p$ system and $6 \%$ for the three-body system.

\begin{acknowledgments}
\noindent
I am thankful to M.Mai for drawing my attention to existence
of the summed up version of the formula. The work was supported by the Czech GACR grant 19-19640S. \hfill
\end{acknowledgments}

\end{document}